\documentclass{mn2e}
\usepackage{epsfig}

\def\kms{\ifmmode {\,\rm km \, s^{-1}}
\else {$\rm km \, s^{-1}$}\fi}
\def\Mpc{\ifmmode {\, h^{-1} \, {\rm Mpc}}
\else {$h^{-1}\,$ Mpc}\fi}

\def\s8{{\sigma_8}}

\def\ltsima{$\; \buildrel < \over \sim \;$}
\def\simlt{\lower.5ex\hbox{\ltsima}} 
\def\gtsima{$\; \buildrel > \over \sim \;$} 
\def\simgt{\lower.5ex\hbox{\gtsima}}

\def\omegam{{\Omega_{\rm m}}}

\def\omegab{{\Omega_{\rm b}}}

\def\omegal{{\Omega_\Lambda}}
 
\def\omegabh2{{\omegab h^2}}

\def\s8m{{\sigma_{8{\rm m}}}}
\def\s8g{{\sigma_{8{\rm g}}}}

\def\hompc{\ifmmode {\,h\,\rm Mpc^{-1}}
\else {$h^{-1}$~Mpc}\fi}
\def\m@th{\mathsurround=0pt }
\def\eqalign#1{\null\,\vcenter{\openup1\jot \m@th
 \ialign{\strut\hfil$\displaystyle{##}$&$\displaystyle{{}##}$\hfil
 \crcr#1\crcr}}\,}
\def\pad{\noalign{\vglue 0.3em}}

\begin{document}

\title[Biasing from the 2dFGRS and the CMB]
{The 2dF Galaxy Redshift Survey: 
The amplitudes of fluctuations in the 2dFGRS and the CMB,
and implications for galaxy biasing}
\author[O.~Lahav et al.]{
%
% Standard MRNAS-style author list for 2dFGRS papers (3/3/2000)
%
\parbox[t]{\textwidth}{
Ofer Lahav$^1$, 
Sarah L. Bridle$^1$, 
Will J. Percival$^{2}$,
John A.\ Peacock$^{2}$,
George Efstathiou$^1$, 
%Ivan K.\ Baldry$^{3}$,
Carlton M.\ Baugh$^3$,
Joss Bland-Hawthorn$^4$,
Terry Bridges$^4$, 
Russell Cannon$^4$, 
Shaun Cole$^3$, 
Matthew Colless$^5$, 
Chris Collins$^6$, 
Warrick Couch$^7$, 
%Nicholas Cross$^$,
Gavin Dalton$^{8}$,
%Kathryn Deeley$^5$, 
Roberto De Propris$^7$,
Simon P.\ Driver$^9$, 
Richard S.\ Ellis$^{10}$, 
Carlos S.\ Frenk$^3$, 
Karl Glazebrook$^{11}$, 
%Edward Hawkins$^{12}$,
Carole Jackson$^5$,
Ian Lewis$^{8}$, 
Stuart Lumsden$^{12}$, 
Steve Maddox$^{13}$,
Darren S. Madgwick$^1$,
Stephen Moody$^1$,
Peder Norberg$^3$,
Bruce A.\ Peterson$^5$, 
%Ian Price$^3$,
%Mark Seaborne$^7$,
Will Sutherland$^{2}$,
%Helen Tadros$^7$, 
Keith Taylor$^{10}$}
\vspace*{6pt}\\ 
$^1$ Institute of Astronomy, University of Cambridge, Madingley Road,
    Cambridge CB3 0HA, UK\\
$^{2}$Institute for Astronomy, University of Edinburgh, Royal Observatory, 
       Blackford Hill, Edinburgh EH9 3HJ, UK\\
$^3$Department of Physics, University of Durham, South Road, 
    Durham DH1 3LE, UK\\ 
$^4$Anglo-Australian Observatory, P.O.\ Box 296, Epping, NSW 2121,
    Australia\\  
$^5$Research School of Astronomy \& Astrophysics, The Australian 
    National University, Weston Creek, ACT 2611, Australia\\
$^6$Astrophysics Research Institute, Liverpool John Moores University,  
    Twelve Quays House, Birkenhead, L14 1LD, UK\\
$^7$Department of Astrophysics, University of New South Wales, Sydney, 
    NSW 2052, Australia\\
$^{8}$Department of Physics, University of Oxford, Keble Road, 
    Oxford OX1 3RH, UK\\
$^9$School of Physics and Astronomy, University of St Andrews, 
    North Haugh, St Andrews, Fife, KY6 9SS, UK\\
$^{10}$Department of Astronomy, California Institute of Technology, 
    Pasadena, CA 91125, USA\\
$^{11}$Department of Physics \& Astronomy, Johns Hopkins University,
       Baltimore, MD 21218-2686, USA\\
$^{12}$Department of Physics, University of Leeds, Woodhouse Lane,
       Leeds, LS2 9JT, UK\\
$^{13}$School of Physics \& Astronomy, University of Nottingham,
       Nottingham NG7 2RD, UK\\
}

\maketitle

\begin{abstract}
We compare the amplitudes of 
fluctuations probed by  
the 2dF Galaxy Redshift Survey
and by the latest measurements of the Cosmic
Microwave Background anisotropies.
By combining the 2dFGRS and 
CMB data  we find the linear-theory rms mass fluctuations in $8 \Mpc$ spheres
to be $\sigma_{8{\rm m}} = 0.73 \pm 0.05$ 
(after marginalization over the matter density
parameter $\omegam$ and three other free parameters).
This normalization is lower than the COBE normalization and 
previous estimates from cluster abundance, but it is
in agreement with some revised 
cluster abundance determinations.
We also estimate  
the scale-independent 
bias parameter of present-epoch $L_s = 1.9L_*$  APM-selected galaxies
to be $b(L_s,z=0) = 1.10 \pm 0.08$ 
on comoving scales of $0.02 < k < 0.15 \hompc$.
If luminosity segregation operates on these scales,
$L_*$ galaxies would 
be almost un-biased, $b(L_*,z=0) \approx 0.96$.
These results are derived by assuming  a flat $\Lambda$CDM Universe, and by  
marginalizing over other free parameters and fixing the 
spectral index $n=1$
and  the optical depth  due to reionization $\tau =0$. 
%the spectral index $n=1$, 
%$\omegam = 1 - \omegal = 0.3$, $\omegab h^2 = 0.02$
%and a Hubble constant $h=0.7$. 
We also study the best fit pair $(\omegam,b)$,
and the robustness of the results to varying 
$n$ and $\tau$. 
%(MORE DETAILS). 
Various modelling corrections  can each change the resulting $b$ 
by 5--15 per cent. 
The results are  compared with 
other independent measurements from the 2dFGRS itself, and from 
the SDSS, cluster abundance and cosmic shear.
\hfill\break
\hfill\break
{\bf Key words}: Cosmology, CMB, galaxies, Statistics 
%
%\begin{keywords}
%Cosmology, CMB, galaxies, Statistics 
%\end{keywords}
\end{abstract}

\strut\vfill\eject
\strut\vfill\eject

%
% fudge needed to stop LaTeX leaving a blank first column
% (yet another reason for using plain TeX)
%

\section{Introduction}\label{intro}

The 2dF Galaxy Redshift Survey (2dFGRS) has now measured 
over 210\,000 galaxy redshifts and is the largest existing galaxy redshift
survey (Colless et al. 2001). 
A sample of this size allows large-scale structure statistics
to be measured with very small random errors. 
Two other 2dFGRS papers, Percival et al. (2001; hereafter P01) and Efstathiou
et al. (2001; hereafter E02) have mainly compared the {\it shape} of the 2dFGRS and CMB
power spectra, and concluded that they are consistent with each other
(see also Tegmark, Hamilton \& Xu 2001).
Here we estimate the {\it amplitudes} 
of the rms fluctuations in mass $\sigma_{8{\rm m}}$ and in galaxies $\sigma_{8{\rm g}}$.
More precisely, we consider the ratio
of galaxy to matter power spectra, and use the ratio of these to
define the bias parameter:
\begin{equation}
b^2 \equiv {P_{\rm {gg}}(k)\over P_{\rm {mm}}(k)}.
\end{equation}
As defined here, $b$ is in principle a function of scale. In practice,
we will measure the average value over the range of wavenumbers
$0.02 < k < 0.15 \hompc$. On these scales, the fluctuations are close
to the linear regime, and there are good reasons (e.g. Benson et al. 2000)
to expect that $b$ should tend to a constant.
In this study,  we will not test the assumption
that the biasing is scale-independent,
but we do allow it to be  function of luminosity
and redshift.

A simultaneous analysis of the constraints placed on cosmological
parameters by different kinds of data is essential because each probe --
e.g. CMB, Type Ia supernovae (SNe Ia), 
redshift surveys, cluster abundance, and peculiar
velocities -- typically constrains a different combination of
parameters (e.g. Bahcall et al. 1999; Bridle et al. 2000, 2001a; E02).
%By performing joint likelihood analyses, one can overcome
%intrinsic degeneracies inherent in any single analysis and so estimate
%fundamental parameters much more accurately. The comparison of
%constraints can also provide a test for the validity of the assumed
%cosmological model or, alternatively, a revised evaluation of the
%systematic errors in one or all of the data sets.  
%Recent papers that
%combine information from several data sets simultaneously include
%Webster et al.\ (1998); Lineweaver (1998); Gawiser \& Silk (1998),
%Bridle et al.\ (1999, 2001), Eisenstein, Hu, \& Tegmark (1999);
%Efstathiou et al.\ (1999); Bahcall et al.\ (1999); Lahav et al (2000), 
%and Efstathiou et al. (2001; E01).
A particular case of joint analysis 
is that of galaxy redshift surveys and the Cosmic Microwave
Background (CMB).
While the CMB probes the fluctuations in matter, 
the  galaxy redshift surveys measure  the perturbations 
in the light distribution  of  particular tracer
(e.g. galaxies of  certain type). 
Therefore, for a fixed set of cosmological  parameters, 
a combination of the two can tell us about the way galaxies
are `biased' relative to the mass fluctuations (e.g. Webster et al. 1998).  

A well-known problem in estimating cosmological parameters is the 
degeneracy of parameters,
and the choice of free parameters.
Here we  consider three classes of parameters:

(i) Parameters that are fixed by theoretical assumptions
or prejudice
(which may be supported by observational evidence).
Here we assume 
a flat Universe (i.e. zero curvature), and no tensor component
in the CMB
(for discussion of the degeneracy with respect to these
parameters see E02).

(ii) The `free parameters' that are 
of interest to address a particular question.
For the 
joint 2dFGRS \& CMB analysis presented here  
we consider five free parameters: the matter 
density parameter $\omegam$,
the linear-theory amplitude of the mass fluctuations $\sigma_{8{\rm m}}$, 
the present-epoch linear biasing parameter 
$b(L_s,z=0)$ (for the survey effective luminosity $L_s \simeq 1.9 L_*$), the 
Hubble constant  $ h \equiv H_0/(100 \kms)$, and
the baryon density parameter $\omega_{\rm b} \equiv \omegab h^2$.
%10 Dec 2001 (OL)
As we are mainly interested in 
combinations of $\sigma_{8{\rm m}}$, $b$ and $\omegam$, 
we shall marginalize
over the remaining  parameters.

(iii) The robustness of the 
results to some `extra parameters', that are 
uncertain. 
Here we consider the optical depth $\tau$ due to reionization 
(see below) and the primordial spectral index $n$.
We use as our canonical values $\tau=0$ and $n=1$, 
but we also quote the results for
 other possibly realistic values, $\tau=(0.05, 0.2)$   
and $n=(0.9,1.1)$.

%At the present paper we use two modes of analysis: 
%(a) 
%we fix  
%parameters  according to 
%the `concordance model' (supported by various observations)
%a flat $\Lambda$CDM universe: 
%spectral index $n=1$, 
%$\omegam = 1 - \omegal = 0.3$ from combined CMB and SNe results 
%(REF),
%$\omegab h^2 = 0.02$ from Big-Bang Nucleosynthesis (REF),
%no tensors, no reionization, 
%and Hubble constant $h=0.7$ from the Cepheid-calibrated distances 
%(Freedman et al. 2000). 
%We then solve only for the galaxy and mass 
%amplitudes $\sigma_{8,g}$ 
%and $\sigma_{8,m}$. 
%
%
%(b) we keep
%$\omegam$, $b(L_s,z=0)$ 
%and   
%$\sigma_{8,m}$ 
%(the matter rms fluctuation  in $8\Mpc$ spheres) as free parameters.
%To derive the most probable biasing parameter
%we then  marginalize over e.g. $\omegam$ and $\s8$
%and over Gaussian probabilities for the Hubble constant
%$h=0.7 \pm 0.07$ (Freedman et al. 2000) and the baryon density parameter 
%$\omegab h^2 = 0.02 \pm 0.002$ (REF).

The outline of this paper is as follows.
In Section 2 we derive $\sigma_{8{\rm g}}$ from the 2dFGRS alone,
taking into account corrections 
for redshift-space distortion and for epoch-dependent and luminosity-dependent
biasing.  
In Section 3 we derive $\sigma_{8{\rm m}}$ from  
the latest CMB data.
In Section 4 we present a  joint analysis 
of 2dFGRS \& CMB.
% where several parameters are kept free,
%and are then marginalized out. 
%(or kept fixed in some cases).
Finally, in Section 5 we compare and contrast our  
measurements with other cosmic probes.

\section{The amplitude of the 2\lowercase{d}FGRS fluctuations}

\subsection{$\bf\sigma^{\bf S}_{\bf 8{\rm g}}$ from the fitted power spectrum}

 An initial estimate of the convolved, redshift-space power spectrum of the
 2dFGRS has already been determined (P01), using the
 Fourier-transform-based technique described by Feldman, Kaiser \& Peacock 
 (1994; hereafter FKP) for a sample of 160\,000 redshifts. 
 On scales $0.02<k<0.15 \hompc$, the data are
 robust and the shape of the power spectrum is not affected by
 redshift-space or non-linear effects, though the amplitude
 is increased by redshift-space distortions (see later).
 We use the resulting power spectrum from P01 in this paper to constrain the 
 amplitude of the fluctuations. 

As explained above, we define the bias parameter as the 
square root of the ratio of the galaxy and mass power spectra on large scales.
We shall assume that the mass power spectrum can be described by a
member of the family of models dominated by Cold Dark Matter (CDM).
Such models traditionally have their normalization described by the
linear-theory value of the rms fractional fluctuations in density
averaged in spheres of  $8 \Mpc$ radius: $\sigma_{8{\rm m}}$.
It is therefore convenient to define a corresponding measure
for the galaxies, $\sigma_{8{\rm g}}$, such that we can express the bias parameter
as
\begin{equation}
b = {\sigma_{8{\rm g}} \over \sigma_{8{\rm m}} }.
\end{equation}
The scale of $8 \Mpc$ was chosen historically 
because $\sigma_{8{\rm g}} \sim 1$ from the optically-selected Lick counts
(Peebles 1980),
so it may seem impossible by definition to produce a linear-theory $\sigma_8$ for galaxies.
In practice, we define $\sigma_{8{\rm g}}$ to be the value 
required
to fit a CDM model to the power-spectrum data on linear scales ($0.02<k<0.15 \hompc$).
From this point of view, one might equally  well specify the normalization
via e.g. $\sigma_{20}$; 
however, the $\sigma_8$ parameter is more familiar in the
context of CDM models. The regions of the power spectrum that generate the
$\sigma_8$ signal are at only slightly higher $k$ than our maximum value,
so no significant uncertainty arises from extrapolation.
A final necessary complication of the notation is that we need to distinguish
between the apparent values 
of $\sigma_{8{\rm g}}$ as measured in redshift space
($\sigma_{8{\rm g}}^S$) and the real-space value that would be measured in the
absence of redshift-space distortions ($\sigma_{8{\rm g}}^R$). It is the latter
value that is required in order to estimate the bias.
 
% The FKP method requires determining 
% \begin{equation}
%   A = \int d^3r \bar{n}^2(r)\,w^2(r),
% \end{equation}
% where $n(r)$ and $w(r)$ are the radial number density 

% ----     Old version ------------------
%
%An initial analysis of
%the power spectrum of the 2dFGRS (Percival et al.\ 2001) yields 68per cent
%confidence limits on the total matter density times the Hubble
%parameter $\Omega_{\rm m} h = 0.20 \pm 0.03$, and the baryon fraction
%$\Omega_{\rm b}/\Omega_{\rm m} = 0.15 \pm 0.07$, assuming
%scale-invariant primordial fluctuations, a $\Lambda$CDM universe,
%and a prior on the Hubble
%constant ($h = 0.7 \pm 10$ per cent).
%Although the $\Lambda$CDM model with
%comparable amounts of dark matter and dark energy is rather esoteric,
%it is remarkable that the 2dFGRS measurement 
%shows such good consistency with the other cosmological probes
%(CMB, SNe, BBN).
%This power spectrum is measured in redshift space, on scales
%$ 0.02 < k < 0.15 \hompc$, 
%over which the data are reliable, linear theory holds, and
%the redshift-space distortion does not alter the shape of the power spectrum
%(see P01).
%

%\footnote{For the 2dFGRS $k$ range used at the present 
%analysis it is actually more meaningful
%to quote the rms fluctuations  on say $20 \Mpc$ spheres.
%For reference, in 
%linear theory 
%$\sigma_{20m}^R = 0.48 \;  \sigma^R_{8m}$  
%for the assumed flat $\Lambda$CDM 
%concordance model with $\omegam=0.3, \omega_b=0.02,h=0.7,n=1$.}.

\begin{figure}
%\centering\epsfig{file=2dFGRS_pk_nov01.ps,width=0.95\hsize}
\centering\epsfig{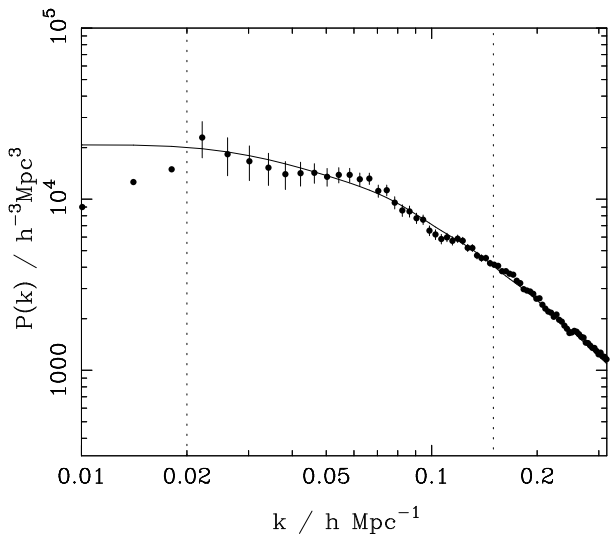}
\caption[]{The observed (i.e. convolved with the window function) 
2dFGRS power spectrum (as derived in P01). 
The  solid line shows  a linear theory $\Lambda$CDM fit 
(convolved with the window function)
with $\omegam h = 0.2, \omegab/\omegam = 0.15, h=0.7, n=1$ 
and best-fitting $\sigma^S_{8{\rm g}}(z_s,L_s) = 0.94$.
Only the range $ 0.02 < k < 0.15 \hompc$ is used 
at the present analysis
(roughly corresponding to CMB harmonics $200 <  \ell < 1500$
in a flat $\omegam = 0.3$ Universe).
The good fit of the linear theory power spectrum 
at $k > 0.15 \hompc$ is due to a conspiracy between the 
non-linear growth and finger-of-god smearing
(integrating over the 
%linear theory model $P(k)$
observed $P(k)$
therefore provides another way of estimating the normalization, giving
$\sigma^S_{8{\rm g}} \simeq 0.95$).
} 
\label{2dF_Pk}
\end{figure}

\begin{figure}
\centering\epsfig{file=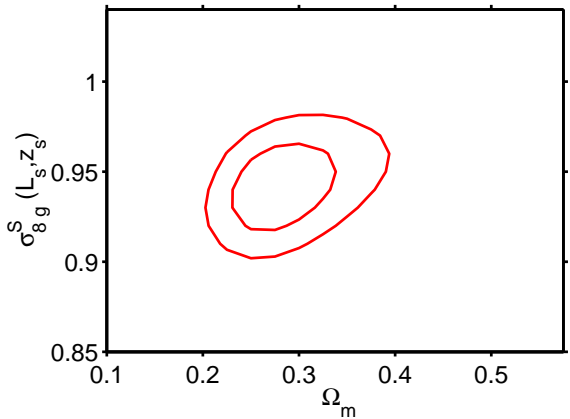,width=0.9\hsize}
\caption[]{The likelihood function of 2dFGRS 
as a function of
of the galaxy fluctuation amplitude in 
redshift space $\sigma^S_{8{\rm g}}(L_s,z_s)$ and the present epoch $\omegam$.
The marginalization over the Hubble constant is done 
with a Gaussian centred at $h=0.7$ and standard deviation of 0.07.
Other parameters are held fixed ($n=1, \omega_{\rm b} = 0.02$).
The contours contain  68 per cent and 95 per cent of the probability.}
\label{2dF_s8_om}
\end{figure}

The 2dFGRS power spectrum (Fig. 1) is fitted in P01 
over the above range in $k$, 
assuming scale-invariant primordial fluctuations and a $\Lambda$CDM cosmology, for 
four free parameters: $\omegam h$, $\omegab/\omegam$, $h$  
and the redshift space $\sigma^S_{8{\rm g}}$ (using the 
transfer function fitting formulae
of Eisenstein \& Hu 1998) . 
Assuming a 
Gaussian prior
on the Hubble constant $h=0.7\pm0.07$ (based on Freedman et al. 2001) 
 the shape of the recovered spectrum
  within the above $k$-range
  was used to yield 68 per cent confidence limits on the 
  shape parameter 
 $\omegam h=0.20 \pm 0.03$, and
 the baryon fraction $\omegab/\omegam=0.15 \pm 0.07$,
 in accordance with the popular `concordance' model\footnote{As 
shown in P01, the likelihood analysis
gives a second (non-standard) solution, 
with   $\omegam h \sim 0.6$, and
the baryon fraction $\omegab/\omegam=0.4$,
which generates baryonic `wiggles'. We ignore this case at the 
present analysis and using the likelihood function over the range
$0.1 < \omegam h < 0.3 $, $0.0 < \omegab/\omegam < 0.4 $, 
$0.4 < h < 0.9 $  
and $0.75 < \sigma^S_{8{\rm g}} < 1.14$.
We also note that even if there are features in the primordial 
power spectrum, they would get washed out by the 2dFGRS 
window function (Elgaroy, Gramann \& Lahav 2002)
}.
Although the $\Lambda$CDM model
with comparable amounts of dark matter and dark energy is rather esoteric,
it is remarkable that the 2dFGRS measurement shows such good consistency
with  other cosmological probes such as CMB, SNe, 
and Big Bang Nucleosynthesis (BBN).

%Cross-sections of the 4-D 2dFGRS likelihood function 
%are shown in Fig.~\ref{cross_4d}.
We find that $\sigma^S_{8{\rm g}}$ depends only weakly on the other
three parameters, with the strongest correlation with $\omegam$.
%This is because  $8 \Mpc$ lies at the small scale end of the 
%$k$ range of 2dFGRS data used for this analysis.
%Therefore, on marginalizing over $\omegam$ the error bars 
%on $\sigma^S_{8g}$ widen slightly.
For fixed 
`concordance model' parameters
$n=1, \omegam = 1 - \omegal = 0.3$, $\omega_{\rm b} = 0.02$
and a Hubble constant $h=0.70$, 
we find that the amplitude of 2dFGRS galaxies
in redshift space is $\sigma_{8{\rm g}}^S (L_s,z_s) = 0.94$
%0.95
(when all other parameters are held fixed, the formal errors are 
unrealistically tiny, only a few percent,
and hence we do not quote them).
%from SLB---------
%is $\sigma_{8,g}^S (L_s) = 0.95^{+0.018}_{-0.008}$.
In the FKP method, the normalization of the power
spectrum depends on the radial number density and weighting function, and a
number of different methods have been suggested for calculating the normalization
(Sutherland et al. 1999) using a random catalogue designed to
 Poisson sample the survey region. P01  tried all of the suggested   
 methods for the 2dFGRS data and found no significant change in the power
 spectrum normalization. Therefore, although this
 calculation remains a potential cause of systematic
 error in the power-spectrum normalization, we shall assume hereafter
that the main uncertainty in the derived bias derives from the
uncertain cosmological model that is needed in order to
connect the galaxy power spectrum with the mass power spectrum from the CMB.

On keeping $\omegam$ free and 
marginalizing over $h$ with a Gaussian prior 
(with $h = 0.7 \pm 10$ per cent)
we obtain
Fig.~\ref{2dF_s8_om}. 
The external constraint on $h$ that we impose translates
to a constraint on $\omegam$ through the 2dFGRS sensitivity
to the matter power spectrum shape, which is roughly 
$\omegam h$. 
On marginalizing over $\omegam$ we find $\sigma^S_{8{\rm g}} =  0.94 \pm 0.02$,
in agreement with the best-fit non-marginalized result\footnote{
We emphasise again that here  $\sigma_{8{\rm g}}$ is
the linear-theory normalization, not the observed non-linear 
$\sigma_{8{\rm g}\rm NL}$. 
For example, the 2dFGRS correlation function of Norberg et al.
(2001a)
can be translated to a non-linear
$\sigma_{8{\rm g}\rm NL}^R(L_*) = 0.87 \pm 0.07$,
at an effective redshift of approximately 0.07.
In practice, nonlinear corrections to $\sigma_8$ are expected to
be relatively small for CDM-like spectra (see Fig. 1).
}.

%\begin{figure}
%\centerline{\vbox{
%\epsfig{file=cl_h07_tau00.ps,width=8.5cm}}}
%\caption[]{A compilation of the latest CMB ${\Delta T \over T}$ data points
%(open circles with error bars)
%against spherical harmonic $l$ (from WTZ).  The line
%shows the predicted angular power spectrum for a $\Lambda$CDM model
%with $n=1$, $\Omega_{\rm m} = 1 -\Omega_\Lambda = 0.3$, $\Omega_{\rm
%b} h^2 = 0.02$ (BBN value), $h=0.7$, 
%COBE normalization 
%$\sigma_{8{\rm m}} =0.90$ (dashed line) and the best-fit to WTZ data 
%$\sigma_{8{\rm m}} =0.77$ (solid line). 
%The stars indicate this best fit model convolved with the 
%experimental window functions.
%A similar model is also
%the best-fit to the shape of the 2dFGRS galaxy power spectrum (Fig. 1).
%[Change y-axis to $T_0^2 \ell(\ell +1) C_{\ell}/(2 \pi)]$.
%}
%\label{dt}
%\end{figure}

\begin{figure}
%\centering\epsfig{file=cmbdata.ps,width=0.95\hsize}
\centering\epsfig{file=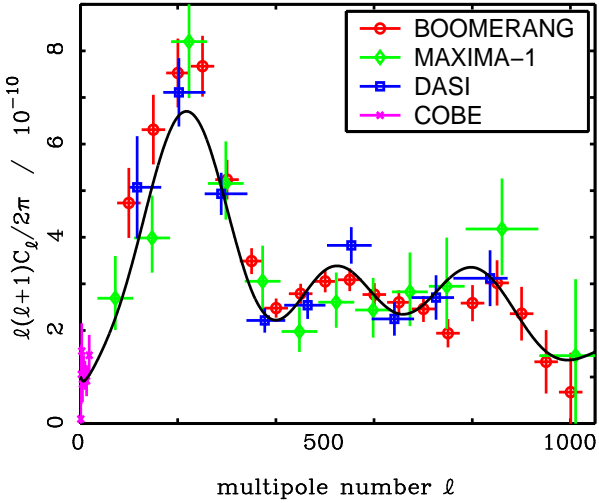,width=0.95\hsize}
\caption[]{A compilation of the latest CMB  data points
from COBE, Boomerang, Maxima and  DASI
against spherical harmonic $\ell$.  The line
shows the predicted angular power spectrum for a $\Lambda$CDM model
with $n=1$, $\omegam = 1 -\Omega_\Lambda = 0.3$, 
$\omega_{\rm b} = 0.02$ (BBN value), $h=0.70$, $\tau =0.0$, 
and the best-fit normalization to the given CMB data points 
$\sigma_{8{\rm m}} =0.83$.
Note that this normalization is lower than the traditional COBE-only
normalization (see Table 1).
A similar model is also
the best fit to the {\it shape} of the 2dFGRS galaxy power spectrum (Fig. 1).
}
\label{dt}
\end{figure}

\subsection {Corrections for redshift and luminosity effects}

In reality, the effective redshift for the P01 analysis is not zero, but 
$z_s \sim 0.17$.
This is higher than the median redshift of 2dFGRS ($z_m \sim 0.11$)
due to the weighting scheme used in estimating the power spectrum.
Similarly, $L_s\simeq 1.9L_*$, rather than the $L_s\simeq L_*$ that
would apply for a flux-limited sample.
We can then derive $\sigma^S_{8{\rm g}}(L_s,z_s)$ directly, but for comparison
with other studies we make further steps of calculating 
$\sigma^R_{8{\rm g}}(L_s,z=0)$ and then  $\sigma^R_{8{\rm g}}(L_*,z=0)$. 
%Then,  to compare the derived galaxy $\sigma_{8g}$  
%with the CMB $\sigma_{8{\rm m}}$ (defined here for redshift zero) we need
%to apply a set of corrections. 
This requires  corrections 
which depend on the nature of galaxy formation 
and on clustering with redshift.
Some of the corrections themselves depend on cosmological parameters,
and our procedure solves for the best-fitting values in a self-consistent way.

We start by evaluating the conversion from redshift space to real space 
at the survey effective redshift $z_s$ 
for galaxies with effective luminosity $L_s$:
\begin{equation}
\sigma_{8{\rm g}}^{R}(L_s, z_s) = 
\sigma_{8{\rm g}}^{S}(L_s,z_s)/K^{1/2}[\beta(L_s, z_s)] \; \;.
\end{equation}
where
\begin{equation}
K[\beta] = 1 + {\frac{2}{3} } \beta + {\frac {1}{5}} \beta^2
\end{equation}
is Kaiser's (1987) factor, 
derived in linear theory and the distant-observer approximation\footnote{More 
precisely, the redshift-space distortion factor depends on
the auto power spectra $P_{{\rm mm}}(k)$ and $P_{\rm {gg}}(k)$ for the mass and the
galaxies, and on the mass--galaxies cross power spectrum $P_{\rm {mg}}(k)$
(Dekel \& Lahav 1999; Pen 1998; Tegmark et al. 2001). The model of
equations (3--5) is only valid for a scale-independent bias factor $b$ that
obeys $P_{\rm{gg}}(k) = b P_{\rm{mg}}(k) = b^2 P_{\rm{mm}}(k)$.}.
The dependence of $\beta$ on redshift can be written as: 
\begin{equation}
\beta(L_s, z_s) \simeq \omegam^{0.6}(z_s)/b(L_s,z_s),
\end{equation}
assuming linear biasing
[for more general biasing schemes see e.g. Dekel \& Lahav (1999) 
and references therein].

The evolution of the matter density parameter with redshift is
\begin{equation}
\omegam(z) = \omegam\; (1+z)^3\; (H/H_0)^{-2}
\end{equation}
with
\begin{equation}
(H/H_0)^2 = 
[\omegam(1+z)^3 +
(1-\omegam-\omegal)(1+z)^2 + \omegal]\;.
\end{equation}

% Nishioka et al - talk at GR16).
The variation of $b(z)=\sigma_{8{\rm g}}(z)/\sigma_{8{\rm m}}(z)$  
with redshift is even more model-dependent.
We assume that the mass fluctuations grow  as
$\sigma_{8{\rm m}}(z) = \sigma_{8{\rm m}}(0) D(z)$,
where $D(z)$ (normalized to 1 at $z=0$) is the growing mode of fluctuations
in linear theory 
[it depends on $\omegam$ and $\omegal$, e.g. Peebles (1980)].

We also assume 
that galaxy clustering weakly evolves over $0< z< 0.2$,
i.e. $ \sigma_{8{\rm g}}(L_s,0) \simeq \sigma_{8{\rm g}}(L_s,z)$.
We shall refer to this simple 
model as the `constant galaxy clustering (CGC) model'.
Simulations suggest 
(e.g. Kauffmann et al. 1999; Blanton et al. 2000; 
Benson et al. 2000;  Somerville et al. 2001) that    
even if the clustering of dark matter halos 
evolves slightly over this range of redshifts,  
galaxy clustering evolves much less. 
Indeed, observationally
there is only a weak evolution of clustering of the overall galaxy population
over the redshift range $0.1 < z < 0.5$ 
(e.g. the CNOC2 survey: Shepherd et al. 2001).
Therefore in our simple CGC model (for any luminosity):
\begin{equation}
b(L_s,z_s) = b(L_s,0)/D(z_s).
\end{equation}

There are of course other possible models for the evolution of
galaxy clustering 
with redshift, e.g.
the galaxy conserving model (Fry 1996). This model  
describes the evolution of bias for test particles
by assuming that they follow the cosmic flow.
It can be written as:
\begin{equation}
b(L_s,z_s) = 1+ [b(L_s,0)-1]/D(z_s).
\end{equation}
More elaborate models exist, such as those based on a merging model
(e.g. Mo \& White 1996; Matarrese et al. 1997; Magliocchetti et al. 1999)
or numerical and semi-analytic models 
(Benson et al. 2000; Somerville et al. 2001).

To estimate the magnitude of these effects we consider 
the 2dFGRS effective redshift $z_s = 0.17$.  
For  a Universe with present-epoch $\omegam=0.3$ and $\omegal=0.7$ we get 
$\omegam(z_s) = 0.41$ and 
$D(z_s) = 0.916$, hence for the 
CGC model with $\omegam=0.3$, $b(L_s,z_s) = 1.09 \; b(L_s,0)$ and 
$\beta(L_s,z_s) = 1.10 \; \beta(L_s,0)$.

On the other hand, we can also relate the amplitude
of galaxy clustering to the  
present-epoch mass fluctuations 
$\sigma_{8{\rm m}}$,
which can be estimated from  the CMB (see below):
\begin{equation}
\sigma_{8{\rm g}}^{R}(L_s,0) = b(L_s,0)\sigma_{8{\rm m}}(0)\;.
\end{equation}
Hence by combining equations (3), (8) and (10) we can solve for
$b(L_s,0)$.

Finally, there is the issue of luminosity-dependent biasing.
Although controversial for some while, this effect has now
been precisely measured by the 2dFGRS (Norberg et al. 2001a,2002);
see also recent results from the 
SDSS (Zehavi et al. 2001). Norberg et al. (2001a) 
found from correlation-function
analysis  that on scales 
$\simlt 10 \Mpc$  
\begin{equation}
b(L,0)/b(L_*,0) = 0.85 + 0.15(L/L_*)\;. 
\end{equation}
If we assume that this relation also applies in the linear regime
probed by our $P(k)$ on scales $0.02 < k < 0.15 \hompc$, then 
the linear biasing factor for $L_*$ galaxies
at redshift zero is 1.14 smaller  
then that for the  2dFGRS galaxies with effective survey luminosity 
$L_s=1.9 L_*$ ($K$-corrected). However, this is a source of uncertainty,
and ultimately it can be answered with the complete 2dFGRS and SDSS
surveys by calculating the power spectra in luminosity bins.

%the fluctuation amplitude is higher for more luminous objects,
%or in terms of relative biasing:
%\begin{equation}
%B(L/L_*) \equiv b(L)/b(L_*) \equiv (r_0/r_0,*)^{\gamma/2} =  0.85 + 0.15 L/L_*\;.
%\end{equation}

The luminosities in equation (11) have been $K$-corrected
and also corrected for passive evolution of
the stellar populations, but the clustering has been measured
at various median redshifts
(for galaxies at the redshift range $ 0.02 < z < 0.28$).
Possible variation of galaxy clustering with redshift is still within 
measurement errors of Norberg et al. (2001a,2002).
For simplicity we shall assume in accord with our CGC model 
that this relation  
is redshift-independent over the redshift range of 2dFGRS ($z\la 0.2$).
We see that 
the effects of redshift-space 
distortion and luminosity bias 
are quite significant, at the level 
of more than $10 $ per cent each.

%Table 1 illustrates the effects of redshift-space 
%distortion and luminosity bias. 
%We see that these corrections are quite significant, at the level 
%of more than $10 $ per cent each.

%\begin{table}
%\label{table}
%\caption{
%Illustrating the effect of 
%various  corrections to the 2dFGRS galaxy
%clustering amplitude $\sigma_{8{\rm g}}$. 
%For the first entry we assumed 
%$n=1, \omegam = 1 - \omegal = 0.3$, $\omega_{\rm b}  = 0.02$
%and $h=0.7$. 
%In the second entry we also assumed
%$\beta(L_s,z_s) = 0.43$ (based on Peacock et al. 2000) and equation (3).
%In the third entry we  assumed a  correction 
%(equation 11) 
%in converting the clustering 
% $L_s=1.9 L_*$ galaxies to $L_*$ galaxies 
%(Norberg et al. 2000a).
%We do not quote formal errors  as they are unrealistically tiny 
%when other parameters are held fixed.
%}
%\center{\vbox{
%\begin{tabular}{@{}lr@{}}
%\begin{tabular}{@{}lrrr}
%\hline
%Data &  $\sigma_{8{\rm g}}$\\
%\hline
%$\sigma_{8{\rm g}}^S(L_s,z_s)$ from 2dFGRS $P(k)$ &  $0.94$\\
%\pad
%$\sigma_{8{\rm g}}^R(L_s,z_s)$ from 2dFGRS $P(k)$ & $0.83$\\
%\pad
%$\sigma_{8{\rm g}}^R(L_*,z_s)$ from 2dFGRS $P(k)$ & $0.73$\\
%\hline
%%$\sigma_{8g}$ from SDSS &  $ ?? \pm ??$\\
%\end{tabular}
%}}
%\end{table}

\bigskip

\section {The CMB data}

The CMB fluctuations are commonly represented by the 
spherical harmonics $C_{\ell}$.
The connection between the harmonic $\ell$ and $k$ is roughly
\begin{equation}
\ell \simeq k\, d_A
\end{equation} 
where for a flat Universe the angular distance to the last scattering surface
is well approximated by (Vittorio \& Silk 1991):  
\begin{equation}
d_A \simeq {\frac{2c}{H_0 \,\Omega_{\rm m}^{0.4}}}\;.  
\end{equation} 
For $\omegam =0.3$ 
the 2dFGRS  range $0.02 < k < 0.15 \hompc$
corresponds 
approximately to $ 200 < \ell < 1500$, which is 
well covered by the recent CMB experiments.
We obtain theoretical CMB power spectra using the
CMBFAST  code (Seljak \& Zaldarriaga 1996).

The latest CMB measurements 
from Boomerang (Netterfield et al.\ 2001, de Bernardis et al.\
2002), Maxima (Lee et al. 2001; Stomper et al.\ 2001) 
and DASI (Halverson et al. 2002; Pryke et al.\ 2002) 
suggest three acoustic peaks.  Parameter fitting
to a $\Lambda$CDM model indicates consistency between the different
experiments, and a best-fit Universe with zero curvature, and an
initial spectrum with spectral index $n  \simeq 1$
(e.g. Wang, Tegmark \& Zaldriaga \ 2001, hereafter WTZ;  E02
and references therein).  Unlike the earlier Boomerang and Maxima
results, the new data also show that the baryon contribution is
consistent with the Big Bang Nucleosynthesis value $\omega_{\rm b} 
\simeq 0.02$ (O'Meara et al.\ 2001).

Various CMB data sets can be combined in different ways (e.g. Jaffe
et al.\ 2001; Lahav et al.\ 2000).  
Here we consider two  compilations of CMB data:

(a) 
a compilation of  COBE (8 points), Boomerang, Maxima and DASI
(hereafter CBMD).
%The 8 COBE points are taken from Tegmark (REF??).
The total number of data points in this compilation is 49,
plotted in Fig.~\ref{dt}.

(b) a compilation of 24 $\Delta T/T $ data points from Wang et al.\
(2001; WTZ), which is based on 105 band-power measurements 
of almost all available CMB experiments (including the
latest Boomerang, Maxima, and DASI data).

Both compilations take into account the calibration errors, which are crucial 
for estimating the amplitude of fluctuations.
For  our compilation (a)
we use a fast method for marginalization 
over calibration and beam uncertainties that assumes a Gaussian 
prior on the calibration and beam corrections (Bridle et al. 2001b).
%These uncertainties are also taken into account in the Wang et al. 
%(2001) compilation.
We apply the usual multi-variate $\chi^2$ procedure 
(e.g. Hancock et al. 1998), 
taking into account the window functions and the covariance matrix
(when available).
Since the Boomerang and Maxima window functions and correlation
matrices are not yet available, we assume that the data points are
uncorrelated and use top-hat window functions (as did WTZ).
This assumption is validated by the fact that we obtain sensible
values of $\chi^2$ for the best-fitting models.

%We start by fixing all the other
%parameters, apart from the amplitude $\sigma_{8{\rm m}}$.
%We assume that CMB fluctuations arise
%from adiabatic initial conditions with Cold Dark Matter and negligible
%tensor component, in a flat Universe with 
%the `concordance' parameters as before:
%$\omegam= 1 -
%\Omega_\Lambda = 0.3$, $n=1$, $h=0.7$ 
%and $\omega_b = 0.02$.  

%This choice is
%motivated by numerous other studies that combined CMB data with other
%cosmological probes (e.g. Bridle et al.\ 2000; Hu et al.\ 2001; Wang
%et al.\ 2001; Section 4 above).  
%Of course, one may keep more free
%parameters, and marginalize over some of them, as done in numerous
%other studies.  
%We note that if more cosmological parameters are
%left free and then marginalized over, the errors  would typically
%be much larger.  

\subsection{CMB-only fits}

We first consider the constraints arising from the CMB data alone.
Table 1 summarizes various estimates for 
$\sigma_{8{\rm m}}$  from the above two new data sets. Note that these
differ from the normalization returned by CMBFAST when only the
COBE data are considered.
This Table  also  illustrates the sensitivity of the results to 
the optical depth to reionization $\tau$ (see below).
We see that differences in data sets
and in assumptions on other parameters can easily lead to uncertainties 
of $\sim 10 $ per cent in the resulting $\sigma_{8{\rm m}}$.
Note that the normalizations derived from WTZ are lower than
those derived from our compilation. This reflects that fact that WTZ
chose to adjust downwards the calibrations of the principal datasets
that we prefer to adopt. We incorporate the calibration uncertainties,
but make no such adjustment.

\begin{table}
\label{table}
\caption{Normalizations for matter fluctuations derived by
fitting to CMB data alone.
In all the entries (unless otherwise stated) other 
parameters are fixed at
$\omegam = 1 - \omegal = 0.3$, $\omega_{\rm b} = 0.02$, $n=1$ and $h=0.7$.
The first three entries 
were derived 
via CMBFAST using the COBE points according to 
the normalization procedure of Bunn \& White (1996).
The other entries were derived by the best fit 
multi-variate $\chi^2$ 
(including the covariance matrix and window functions) 
for the WTZ and CBMD data points
(the goodness of fit is e.g.
$\chi^2 = 35$ for 24 points in the fourth entry).
Quoted error bars are 1-sigma.
In some cases
formal errors are not quoted as they are unrealistically tiny (few percent) 
when other parameters are held fixed.
Note that both the WTZ and CBMD compilations give normalization lower
than the  COBE-only normalization.
}
\center{\vbox{
%\begin{tabular}{@{}lrrr}
\begin{tabular}{@{}lr@{}}
\hline
Data &  $\sigma_{8{\rm m}}$\\
\hline
COBE ($\tau=0$), &  $0.90$\\
\pad
COBE ($\tau=0.05$), &  $0.93$\\
\pad
COBE ($\tau=0.20$), &  $0.98$\\
\pad
WTZ ($\tau=0$),  & $0.77$\\
\pad
WTZ ($\tau=0.05$),  & $0.80$\\
\pad
WTZ ($\tau=0.20$),  & $0.92$\\
\pad
CBMD ($\tau=0$) & $0.83$\\
\pad
CBMD($\tau=0$, {\rm marg. over} $h= 0.7 \pm 0.07$ ) & $0.71 \pm 0.07$\\
\pad
CBMD ($\tau=0$, {\rm marg. over} $h= 0.7 \pm 0.07$ 
\&  $\omegam$) & $0.68 \pm 0.07$\\
\pad
CBMD ($\tau=0$) + 2dF, {\rm marg. over} $h$, 
$\omegam$, $\omega_b$ \& $b$ & $0.73 \pm 0.05$\\
\hline
\end{tabular}
}}
\end{table}

%\subsection {$\sigma_{8{\rm m}} $ from the C+B+M+D compilation}
%Fixing $h=0.70$ and $\Omega_{\rm m}=0.3$ and $\Omega_{\rm b} h^2=0.02$ 
%and using the compilation COBE+Boomerang+Maxima+DASI, 
%we find $\sigma_{8{\rm m}}= 0.832 \pm 0.024$.

\begin{figure}
\centering\epsfig{file=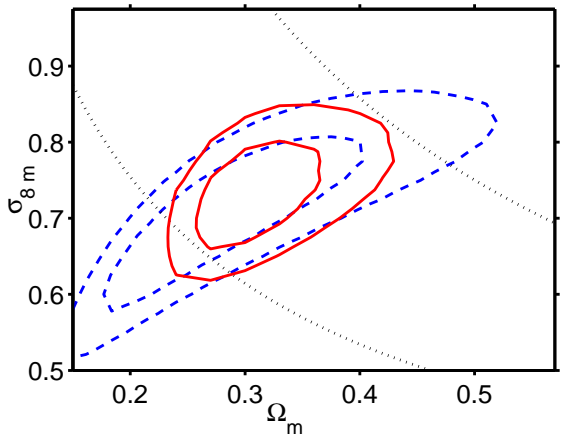,width=0.9\hsize}
\caption[]{The likelihood function of CMB alone  (dashed lines) in terms
of the mass fluctuation amplitude $\sigma_{8{\rm m}}$ and the present 
epoch $\omegam$.
The marginalization over the Hubble constant is done 
with a Gaussian centred at $h=0.7$ and standard deviation of 0.07.
Other parameters are held fixed ($n=1, \omega_{\rm b} = 0.02, \tau=0.0$).
The contours are for (two-parameter) 68 per cent and 95 per cent confidence intervals.
The solid lines show the contours (68 per cent and  95 per cent) 
for the joint 2dFGRS\&CMB analysis,
after marginalization over $h$, $b(L_s,0)$ and $\omega_{\rm b}$.
Other parameters are held fixed ($n=1, \tau=0.0$).
Note that the contours of 2dFGRS \& CMB are much tighter than 
when using CMB alone. 
Two recent extreme cluster abundance determinations 
are overlayed as  the upper
dotted line (Pierpaoli et al. 2001)
and the lower dotted line (Viana et al. 2002).
}
\label{cmb_s8_om}
\end{figure}

Fig.~\ref{cmb_s8_om} (dashed lines)
shows the likelihood as a function of
$(\omegam, \sigma_{8{\rm m}})$ after 
marginalization over the Hubble constant is done 
with a Gaussian with $h=0.7 \pm 0.07$, 
while keeping
other parameters fixed ($n=1, \omega_{\rm b}  = 0.02, \tau=0.0$).
We note that for a fixed $\omegam =0.3$ on this diagram the resulting 
$\sigma_{8{\rm m}} \sim 0.7$ is lower that the value we obtained above
($\sigma_{8{\rm m}} \sim 0.8$) when fixing the Hubble constant 
$h=0.7$ and other parameters.
This illustrates the sensitivity of the results 
from the CMB alone to the Hubble constant. 
The external constraint on $h$ we have imposed cuts off the 
contours at low and high $\omegam$. This is due to the
constraint on $\omegam h^2$ that exists from CMB data:
a constraint on $h$ thus translates to a constraint on 
$\omegam$.
Completing the marginalization over $\omegam$ we find
$\sigma_{8{\rm m}}=0.68 \pm 0.07$.
% From Sarah 4 Dec  (in response to WS)
Note that since we assume that the Universe is flat, there are additional
constraints on our free parameters that come from the \emph{position} of
the first acoustic peak that make our error bars slightly smaller than
studies that marginalize over the curvature of the Universe as well.
%From Fig.~\ref{cmb_s8_om} 
%we can see that these lower values
%of $\sigma_8$ are contributed to by the lower values of 
%$\omegam$ allowed by the data. 
We overlay in Fig.~\ref{cmb_s8_om} the constraints 
from cluster abundance obtained recently by various authors.
%$\sigma_{8{\rm m}}\; \omegam^{0.6}  \simeq (0.495^{+0.034}_{-0.037})$
%obtained 
%by Pierpaoli, Scott and White (2001) and note that
The cluster abundance 
 constraint is fortunately orthogonal to the CMB constraint,
but the spread in normalization values is quite large.
It is interesting that some of the latest estimates are in good agreement
with our estimates from the CMB and 2dFGRS+CMB
(see further discussion below).

%this is  higher than the CMB constraint (cf. E01), 
%but interestingly orthogonal.
%

\section {Combining 2\lowercase{d}FGRS \& CMB} 

When combining 2dFGRS and CMB data
the parameterisation for the log-likelihoods is
in five parameters:

\begin{equation}
\eqalign{
{\ln {\mathcal L}}_{\rm tot} &= \ln {\mathcal L}_{\rm 2dFGRS} [\omegam, h, 
\omega_{\rm b}, \sigma_{8{\rm m}}, b(L_s,0)] \cr
& + \ln  {\mathcal L}_{\rm CMB} [\omegam, h, \omega_{\rm b}, \sigma_{8{\rm m}}],
}
\end{equation}
where
${\mathcal L}_{\rm 2dFGRS}$ and
${\mathcal L}_{\rm CMB}$ 
are the likelihood functions for 2dFGRS and the CMB.

The 2dFGRS likelihood function takes into account the redshift-space distortions, 
the CGC biasing scheme, and the redshift evolution of $\omegam(z)$.
Here we use our compilation of 49 CMB data points (shown in Fig. 3).
Other parameters are held fixed 
($n=1, \tau=0$).

Fig. ~\ref{cmb_s8_om} (solid lines) shows 
the 2dFGRS+CMB likelihood as a function of
$(\omegam, \sigma_{8{\rm m}})$, after marginalization over
$h, b(L_s,0)$ and $\omega_{\rm b}$.
The peak of the distribution is consistent with the result 
for the CMB alone (shown 
by the dashed lines in Fig. ~\ref{cmb_s8_om}),
but we see that the contours are tighter due
to the addition of the 2dFGRS data. 
Further marginalization over $\omegam$  
gives  $\sigma_{8{\rm m}} = 0.73 \pm 0.05$.
The  importance of adding the shape information of the 2dFGRS power spectrum
is that it requires no external prior for 
$h$ and $\omega_{\rm b}$,
unlike deriving $\sigma_{8{\rm m}}$ from CMB alone (Table 1).  
Our result is very similar to the value
$\sigma_{8{\rm m}} \simeq 0.72 $
derived in E02 
using the WTZ data set and after marginalizing over 
the raw 2dFGRS amplitude of the power spectrum and other parameters.

\begin{figure}
\centering\epsfig{file=
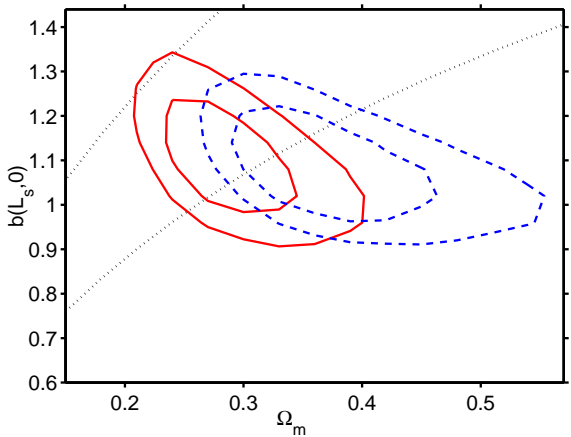,width=0.9\hsize}
\caption{
The result of a joint likelihood 2dFGRS+CMB (solid lines).
The marginalization (without any external  priors) 
is over $h, \omega_{\rm b}$ and 
$\sigma_{8{\rm m}}$. Other parameters are held fixed 
($n=1, \tau = 0$).
The contours are for (two-parameter) 68 per cent and 95 per cent confidence intervals.
The  dotted  lines represent  the 1-sigma envelope for
$\beta(L_s,0)$,
based on  $\beta(L_s,z_s) = 0.43 \pm 0.07$ from Peacock et al (2000)
and the CGC model.
The result of a joint likelihood 2dFGRS+CMB
for $n=0.9$ is marked by the dashed contours (68 per cent and 95 per cent).
}
\label{b_om_n1}
\end{figure}

To study the biasing parameter
we marginalize the 2dFGRS likelihood 
over $h, \omega_{\rm b}$ and 
$\sigma_{8{\rm m}}$. 
Other parameters are held fixed 
($n=1, \tau=0$).
The resulting likelihood as a function of 
$[\omegam, b(L_s,0)]$ is shown (by solid contours) in Fig.~\ref{b_om_n1}.
Further marginalizing over $\omegam$ gives
$b(L_s,0)=1.10 \pm 0.08$ (1-sigma).
With Fry's biasing scheme (equation 9) $b(L_s,0)$ is increased by $8 $ per cent.

The effect of changing the spectral index to $n=0.9$
is shown (by dashed lines) in 
Fig.~\ref{b_om_n1} 
(with $\tau = 0$).
Results for $n=0.9$ and $n=1.1$ with further marginalization 
over $\omegam$ 
are given in Table 2, showing that  
$b(L_s,0)$
is slightly down and up respectively relative to the standard $n=1$ case. 
%Further marginalizing over 
%$b(L_s,0)=0.96 \pm 0.09$ (1-sigma), i.e. a 4 per cent drop 
%compared with $n=1$.
%Conversely, $n=1.1$ gives 
%$b(L_s,0)= 1.07 \pm 0.09$ (1-sigma).
We see that when we fit CMB data over a wide range of $\ell$,
the effect of changing $n$ is small.
This is in contrast with the large variation of fitting the normalization 
with COBE only, where for the concordance model
$\sigma_{8{\rm m}} = (0.72, 0.90, 1.13)$ for
$n=(0.9, 1.0, 1.1)$, respectively.

We also tested sensitivity to the optical depth $\tau$.
Recent important constraints come from the spectra of SDSS quasars,
suggesting $\tau \ga  0.03-0.04$ (Becker et al. 2001; Fan et al. 2002). 
For fixed $n=1, \omega_{\rm b} = 0.02$, and marginalization 
over $\omegam, \sigma_{8{\rm m}}$ and $h$ we get 
$b (L_s,0) = 1.06 \pm 0.09$ for $\tau=0.05$, 
i.e. lower by 
4 per cent compared with  the case of
$\tau=0.0$. Note that setting $\omega_{\rm b} =0.02$ 
or marginalizing over it makes little 
difference to $b(L_s,0)$.
The  effect of the optical depth 
is indeed expected to increase $\sigma_{8{\rm m}}$ 
by a factor $\exp(\tau)$,
and hence to decrease $b$ by that factor, about 5 per cent in the case
of $\tau=0.05$ (corresponding to redshift of reionization 
$z_r \simeq 8$ for the concordance model parameters; 
e.g. Griffiths \& Liddle 2001).

Other possible extra physical parameters may also slightly affect our result.
For example,  a neutrino with mass of 0.1~eV
(e.g. Hu, Eisenstein \& Tegmark 2001; 
Gawiser 2001) would reduce $\sigma_{8{\rm m}}$ by a few  per cent.

Finally, to translate the biasing parameter from $L_s$ to
e.g. $L_*$ galaxies one can either assume
(somewhat ad-hoc) no luminosity segregation on large 
scales, or divide by the factor 1.14 (equation 11) that applies
on small scales.
E.g. using the fully marginalized result 
$b(L_s,0) \simeq 1.10$ we get $b(L_*,0) \simeq 0.96$,
i.e. a slight anti-bias.
Overall, our results can be described by the following formula:
\begin{equation}
b(L_*,z=0) = (0.96 \pm 0.08)\, \exp[-\tau + 0.5(n-1)].
\end{equation}

%\subsection{The bias parameter from both}

%With $h=0.72$ but marginalizing over $\Omega_{\rm m}$
%and including the redshift space correction (taking into account
%the variation of $\Omega_{\rm m}$ with redshift) but assuming
%that bias is constant with redshift we find
%$b(L_s,0)=1.12\pm 0.16$.
%Using a bias scheme in which the galaxy clustering amplitude is
%constant with redshift we obtain $b(L_s,0)=1.07\pm 0.15$.
%With Fry's biasing (equation 11) $b(L_s,0)=1.15\pm 0.15$.
%Now marginalizing over $h$ using a Gaussian prior [!!! still top 
%hat at the moment !!! currently over $0.62<h<0.82$] we get the
%plot in Fig.~\ref{b_om}.
%Marginalizing over $\Omega_{\rm m}$
%$b(L_s,0)=1.04\pm 0.19$.
%the joint CMB+2dFGRS  with  
%top-hat prior on $0.63 < h < 0.77$,
%the best fit parameters are: 
%$\sigma_8=0.73, h = 0.66, 
%\omegam=0.3$ and $b(L_s,0) = 0.99$ (add errors).
%
%
%

%{\bf Based On Sarah's plots (11 Oct 01)}

\begin{table}
\label{table}
\caption{
The biasing parameter $b(L_s,z=0)$ from the full
Maximum Likelihood solution (equation 14), and marginalization 
over ($h,\omega_{\rm b}, \sigma_{8{\rm m}}, \omegam)$ without any external priors
(apart from the third entry, where $\omega_{\rm b}=0.02$).
}
\center{\vbox{
%\begin{tabular}{@{}lrrr}
\begin{tabular}{@{}lr@{}}
\hline
Data & $b(L_s,0)$\\
\hline
2dFGRS+CBMD ($n=1.0, \tau = 0$) &  $1.10 \pm 0.08$\\
\pad
2dFGRS+CBMD ($n=0.9, \tau = 0$) &   $1.08 \pm 0.09$\\
\pad
2dFGRS+CBMD ($n=1.1, \tau = 0$) &   $1.15 \pm 0.09$\\
\pad
2dFGRS+CBMD ($n=1.0, \tau = 0.05, \omega_{\rm b}=0.02$) &   $1.06 \pm 0.09$\\
\hline
\end{tabular}
}}
\end{table}

\section {Comparison with other measurements }

\subsection{Other estimates of 2dFGRS amplitude of fluctuations}

An independent  measurement from 2dFGRS comes from 
redshift-space distortions on scales $ < 10 \Mpc$ (Peacock et al. 2000).
This gives 
$\beta(L_s,z_s) = 0.43 \pm 0.07$.
In Fig. ~\ref{b_om_n1} we show this constraint, after translating it to
$\beta(L_s,z=0)$ via the CGC model.  
We see consistency with our present analysis at the level of 1-sigma.
%For $\omegam = 0.3$ 
%and our derived $b(L_s,z=0) = 1.00$ 
%we get a somewhat lower value 
%$\beta(L_s,z_s) \simeq  0.53 \pm ??$ (CHECK). 
Using the full likelihood function 
in the $(b, \omegam$) plane (Fig. ~\ref{b_om_n1} ) we derive a slightly 
larger (but consistent) value,
$\beta(L_s,z_s) \simeq  0.48 \pm 0.06$.

A study of the bispectrum of the 2dFGRS (Verde et al. 2001) 
on smaller scales ($0.1 < k < 0.5 \hompc$) sets  constraints 
on deviations from linear biasing, and it gives a best-fit solution 
consistent with  linear biasing of unity.
The  agreement with the result of the present paper is impressive,
given that the methods used are entirely different.
In fact, by matching the two results one can get constraints 
on e.g.  $\tau \la  0.2 $.

\subsection {Comparison with other independent measurements}

\subsubsection {SDSS}

Maximum Likelihood analysis of the early Sloan Digital Sky Survey (SDSS) 
by Szalay et al. (2001) finds  
from the projected distribution of galaxies in the magnitude bin 
$ 20 < r^* < 21 $ (median redshift $z_m = 0.33$)  
% or z = 0.25 ??
a shape parameter $\Gamma \simeq \omegam h = 0.183 \pm 0.04$ 
and a linear real-space
$\sigma^R_{8{\rm g}}(z=0) = 0.785 \pm 0.053$ (1-sigma errors), assuming a flat 
$\Lambda$CDM model with $\omegam = 1 - \omegal = 0.3$, 
for the case of  no evolution of galaxy clustering, 
equivalent to  our CGC model
(see also Dodelson et al. 2001). 
To convert the SDSS $r^*$ magnitude we use models similar to those
given in Norberg et al. (2001b), where we find that for $z_m \simeq 0.33$,
$b_{\rm J} \simeq r^* + 1$ (for the mix of galaxy populations). 
Hence at that redshift $r^* = 20 $ corresponds to 
absolute $M_{b_{\rm J}} \simeq -19.4$ (in a flat $\omegam = 0.3$ Universe),
which with appropriate $K$ and evolution correction gives a rest-frame
$M_{b_{\rm J}} \simeq -19.6$. This is in fact very close to  
$L_*$ of the 2dFGRS.
Hence the derived SDSS value,
$\sigma^R_{8{\rm g}}(L_*, z=0) = 0.785$, is 
in accord with the real-space values of we get from 2dFGRS.

%Detailed comparison is subject 
%assumptions of luminosity evolution across the different apparent-magnitude 
%bins. 
 
%a shape parameter $\Gamma \simeq \omegam h = 0.188 \pm 0.04$ 
%and the linear real-space
%$\sigma^R_{8g}(z=0) = 0.915 \pm 0.06$ (1-sigma errors), assuming a flat 
%$\Lambda$CDM model with $\omegam = 1 - \omegal = 0.3$. 
%Their result 
%assumes that  the evolution of the galaxy clustering 
%follows the clustering evolution of the mass, which 
%is different from our CGC assumption. 

\subsubsection {Cluster abundance}

A popular method for constraining $\sigma_{8{\rm m}}$ and $\omegam$ 
on scales of $\sim 10 \Mpc$ is based on
the number density  of rich galaxy clusters.
%(e.g. Eke et al. 1998; Viana \& Liddle 1999).
Four recent analyses span a wide range of values, 
but interestingly they are all orthogonal to our CMB 
and 2dF constraints 
(Fig. 4) .

Pierpaoli, Scott \& White (2001) derived a high value,
while Seljak (2001), 
Reiprich \& Boehringer (2002), 
and Viana, Nichol \& Liddle (2002) found lower values: 
\begin{equation}
\eqalign{
\sigma_{8{\rm m}} &\simeq  0.50\,\omegam^{-0.6}\cr
\sigma_{8{\rm m}} &\simeq  0.44\,\omegam^{-0.44}\cr
\sigma_{8{\rm m}} &\simeq  0.43\,\omegam^{-0.38}\cr
\sigma_{8{\rm m}} &\simeq  0.38\,\omegam^{-0.48+0.27 \omegam}\cr
}
\end{equation}
respectively. 
For $\omegam=0.3$ these results correspond to 
$\sigma_{8{\rm m}} \simeq  1.02; 0.75; 0.68; 0.61$ respectively
(with typical errors of 10 per cent). 
The high  value agrees with numerous earlier studies by Eke et al. 
(1998) and others which were based on temperature functions, 
and it remains to be understood why the recent values 
are so low.
The discrepancy between the different estimates is in part due to differences
in the assumed mass-temperature relation.
The cluster physics still needs to be better understood before
we can conclude which of the above results is more plausible.
We see in Fig. 4 that the lower cluster abundance 
results are actually in good agreement 
with our value from the 2dFGRS+CMB, $\sigma_{8{\rm m}} \simeq 0.73 \pm 0.05$.

%of X-ray selected clusters
%with improved modelling  gives
%$\sigma_{8{\rm m}}\; \omegam^{0.6}  \simeq (0.495^{+0.034}_{-0.037})$
%for $\Gamma \simeq 0.23$.
%As already mentioned, we see in 
%Fig.~\ref{cmb_s8_om} that this constraint is almost orthogonal to the 
%CMB constraint, but with poor overlap.
%Specifically for the  concordance value  $\omegam = 0.3$ it gives  
%$\sigma_{8{\rm m}} \simeq 1.0$ compared with $\sigma_{8{\rm m}} \simeq 0.7$ 
%from the CMB. 
%The discrepancy might be due to the difference in 
%scales on which the $\sigma_{8{\rm m}}$ are determined 
%(cluster abundance on $\sim 10 \Mpc$, CMB on $\sim 20-50  \Mpc$) 
%and different systematic effects in the two data sets.

%THE NEW PAPERS ...

\subsubsection {Cosmic shear}

The measurements of weak gravitational lensing (cosmic shear) 
are sensitive to the amplitude of the matter
power spectrum on mildly non-linear scales. Van Waerbeke et al.
(2001), 
Rhodes, Refregier \& Groth (2001) and 
and Bacon et al. (2002)  find respectively
\begin{equation}
\eqalign{
\sigma_{8{\rm m}} &\simeq  0.43\,\omegam^{-0.6}\cr
\sigma_{8{\rm m}} &\simeq  0.51\,\omegam^{-0.48}\cr
\sigma_{8{\rm m}} &\simeq  0.43\,\omegam^{-0.68}\cr
}
\end{equation}
(with errors of about 20 per cent).
These estimates  are higher than the $\sigma_8{\rm m}$  value
that we obtain from 2dFGRS+CMB, but note the large error bars
in this recently developed  method.    

%for the  concordance value  $\omegam = 0.3$ it gives  
%$\sigma_{8{\rm m}} \simeq 0.9$,
%somewhat lower than the value from cluster abundance
%and above the CMB values.

\section {Discussion} 

We have combined in this paper the latest 2dFGRS and CMB data.
The first main  result of this joint analysis is 
the normalization of the mass fluctuations, 
$\sigma_{8{\rm m}} = 0.73 \pm 0.05$.
This normalization is lower than the COBE normalization and 
previous estimates from cluster abundance, but it is
actually in agreement with recently revised cluster abundance normalization.
The results from cosmic shear are still somewhat higher,
but with larger error bars. 
 
The second result is for the biasing parameter for
optically-selected $L_s$ galaxies,
$b(L_s,0) = 1.10 \pm 0.08$, which
is just consistent with  no biasing (`light traces mass') on scales
of tens of Mpc.
When translated to $L_*$ via a correction valid for  small 
scales we get a slight anti-bias,
$b(L_*,0) \simeq 0.96$.
Although biasing was commonly neglected until the early 1980s,
it has become evident that on scales $\la 10 \Mpc$  
different galaxy populations exhibit 
different clustering amplitudes, the so-called
morphology-density relation (e.g. Dressler 1980; Hermit et al. 1996; 
Norberg et al. 2002).
Biasing on small scales is also predicted in the simulations
of hierarchical clustering from CDM initial conditions 
(e.g. Benson et al. 2000).
It is important therefore to pay attention to the scale 
on which biasing operates. 
Our result of linear biasing of unity on scales ($\ga  10 \Mpc$)
is actually in agreement with predictions of simulations
(e.g. Blanton et al. 2000 Benson et al. 2000; Somerville et al. 2001).
It was also demonstrated  by Fry (1996)
that even if biasing was larger than unity at high redshift, 
it would converge towards unity at late epochs (see equation 9).

We note that in deriving these results from the 2dFGRS and the CMB,  
we have had to consider various corrections  
due to astrophysical and cosmological effects:

\begin{itemize}

\item redshift-space distortions 
cause the amplitude in redshift space 
to be  $\sim 15 $ per cent larger than that
in real space.

\item The evolution of biasing with redshift 
 (for our simple constant galaxy clustering  model) gives a biasing that is 
 $\sim 10 $ per cent  higher at $z_s = 0.17$ than at redshift zero. 

\item If luminosity-dependent biasing also holds on large scales
  then  
  the biasing parameter $b(L_s=1.9 L_*)$ 
  is $\sim 15$ per cent higher than that of $L_*$ galaxies. 

\item On the CMB side, an optical depth $\tau = 0.05$ due to reionization
  reduces the derived biasing parameter $b$ by $\sim 5 $ per cent.
  Changing the spectral index from $n=1$ to $n=0.9$ 
  (for both the CMB and 2dFGRS)
  also reduces $b$ by $\sim 5 $ per cent.  

\end{itemize} 

While we included these corrections in our analysis 
we note that they are
model dependent, and these theoretical uncertainties combined
may account for $\sim 5-10 $ per cent uncertainty over and above the 
statistical random errors.

It may well be that in the future the
cosmological parameters
will be fixed by CMB, SNe etc.
Then, for fixed reasonable cosmological parameters,
one can use redshift surveys to study biasing, evolution, etc.
This paper is a modest illustration of this approach.
Future work along these lines will include
exploring non-linear biasing models 
(e.g. Dekel \& Lahav 1999; Sigad, Branchini \& Dekel  2001; Verde et al. 2001)
per spectral type 
(Madgwick et al 2002; Norberg et al. 2002; 
Hawkins et al. (2001, in preparation)
and the detailed variation of other galaxy properties with 
local mass density.

\bigskip
\bigskip

%\end{document}

%\section {Possible plots}

%* 2dFGRS $P(k)$ and CMB $C_{\ell}$ side by side,
%with models (predicted by each, and joint 
%fits). Indicate translation from $k$ to $l$ on each.
%* $\omegab/\omegam$ vs. $\omegam h$ for : 2dFGRS; CMB; 2dFGRS+CMB

\section*{ACKNOWLEDGMENTS} 

The 2dF Galaxy Redshift Survey was made possible through
the dedicated efforts of the staff of the Anglo-Australian Observatory,
both in creating the 2dF instrument and in supporting it on the telescope.
We thank Oystein Elgaroy, Andrew Firth and Jerry Ostriker 
for helpful discussions.


\begin{thebibliography}{99}


\bibitem{Bacon}
Bacon D.J., Massey R.J., Refregier A.R., Ellis, R.S. 
2002, MNRAS, submitted, astro-ph/0203134


%David Bacon (1), Alexandre Refregier (1), Richard Ellis
%astro-ph/0003008 Accepted for publication in MNRAS


\bibitem{Bahcall}
Bahcall N.A., Ostriker J.P., Perlmutter S., Steinhardt P.J., 1999, 
Science, 284, 148

\bibitem{}
Becker R.H., et al.,  2001, ApJ, accepted, astro-ph/0108097


\bibitem{}
Benson A.J., Cole S., Frenk C.S., Baugh C.M., 
Lacey C.G., 2000, MNRAS, 311, 793


\bibitem{Blanton}
Blanton M., Cen R., Ostriker J.P., Strauss M.A., Tegmark M., 2000.
ApJ, 531, 1

\bibitem{bridle99} 
Bridle S.L., Eke V.R.  Lahav O.,
Lasenby A.N., Hobson M.P., Cole S., Frenk C.S.,
Henry J.P., 1999,
MNRAS, 310, 565 
%(astro-ph/9903472)

\bibitem{bridle01}
 Bridle S.L., Zehavi I., Dekel A., Lahav O., Hobson M.P., 
Lasenby A.N., 2001a, MNRAS, 321, 333


\bibitem{bridle01}
Bridle S.L., Crittenden R., 
Melchiorri A., Hobson M.P., Kneissl R., Lasenby A., 
2001b, MNRAS, submitted, astro-ph/0112114

\bibitem{}
Bunn E.F.,  White M., 1997, ApJ, 480, 6

\bibitem{}
Colless M. \& the 2dFGRS team,  2001, MNRAS, 328, 1039 

\bibitem{}
de Bernardis P., et al., 2002, ApJ, 564, 559
%2001, astro-ph/0105296


\bibitem{}
Dekel A., Lahav O., 1999, ApJ, 520, 24


\bibitem{dk} 
Dodelson S. \& the SDSS team, 2001, ApJ, submitted, astro-ph/0107421

\bibitem{}
Dressler A., 1980, ApJ, 236, 351



%\bibitem{efstathioublhe98} 
%	Efstathiou G., Bridle S.~L., Lasenby A.~N., Hobson M.~P.,
%	Ellis R.~S. 1999, MNRAS, 303, L47


\bibitem{}	
Efstathiou G. \& the 2dFGRS team, 2002, 
%submitted to MNRAS, 
%astro-ph/0109152
MNRAS, 330, 29 (E02)

\bibitem{} 
Eisenstein D.J., Hu W., 1998, ApJ, 496, 605

%\bibitem{eht98}
%Eisenstein D.J., Hu W., Tegmark M., 1999, ApJ, 518, 2




\bibitem{} 
Eke V.R., Cole S., Frank C.S., Henry P.J., 1998, MNRAS, 298, 1145

\bibitem{} 
Elgaroy O., Gramann M., Lahav O., 2002,
MNRAS, 333, 93
%MNRAS, in press, astro-ph/0111208

\bibitem{}	
Fan, X., et al.,  
2002, AJ, 123, 1247
%2001, astro-ph/0111184

\bibitem{}
Feldman H.A., Kaiser N., Peacock J.A., 1994, ApJ, 426, 23   



\bibitem{}
Freedman W.L., et al., 2001, ApJ, 553, 47 

\bibitem{}
Fry J.N., 1996, ApJ, 461, L65

\bibitem{}
Gawiser E.,  2000, Proceedings of PASCOS99 Conference, Lake Tahoe, CA 1999,
astro-ph/0005475


%\bibitem{gs98}
%Gawiser E., Silk J., 1998, Science, 280, 1405

\bibitem{}
Griffiths, L., Liddle A., 2001, 
MNRAS, 324, 769
%astro-ph/0101149

\bibitem{}
Halverson N.W., et al., 2002, ApJ, 568, 38 
%2001, astro-ph/0104489

\bibitem{}
Hancock S., Rocha G., Lasenby A.N., Gutierrez C.M., 1998, MNRAS, 294, L1

%\bibitem{}
%Hawkins E. \& the 2dFGRS team, 2001, in preparation

\bibitem{}
Hermit S., Santiago B.X., Lahav O.,  
Strauss M.A., Davis M., Dressler A.,
Huchra J.P., 1996, MNRAS, 283, 709 


\bibitem{}
Hu W., Eisenstein D.J.,  Tegmark M.,  1998, Phys. Rev. Lett.,
80, 5255


\bibitem{}
Jaffe A. et al., 2001,
Phys. Rev. Lett., 86, 3475

\bibitem{}
Kaiser N., 1987,  MNRAS, 227, 1

\bibitem{}
Kauffmann G., Colberg J.M., Diaferio A., White
S.D.M., 1999, MNRAS, 303, 188

\bibitem{}
Lahav O., Bridle S.L., Hobson M.P., Lasenby A.L., Sodr\'e L., 
2000, MNRAS, 315, 45L

\bibitem{}
Lee A.T. et al.,  2001, ApJ, 561, L1
%astro-ph/0104459


%\bibitem{}
%Lineweaver C. H., 1998, ApJ, 505, L69

\bibitem{}
Madgwick D.S. \& the 2dFGRS team, 
MNRAS, 333, 133
%2001, astro-ph/0107197

\bibitem{}
Magliocchetti M., Bagla J.,  Maddox S.J.,  Lahav O.  1999,
MNRAS, 314, 546 
%(astro-ph/9902260).

\bibitem{}
Matarrese S., Coles P., Lucchin F., Moscardini L.,  
1997, MNRAS, 286, 95



\bibitem{}
Mo H.J.,  White S.D.M., 1996, 
MNRAS, 282, 347 

\bibitem[]{}
Netterfield C.B. et al., 2001, ApJ, accepted, astro-ph/0104460



\bibitem{}
Norberg P. \& the 2dFGRS team, 2001a, MNRAS, 328, 64
%astro-ph/0105500
%b(L) 

\bibitem{}
Norberg P. \& the 2dFGRS team, 2001b, MNRAS, submitted, astro-ph/0111011
%LF and selection function

\bibitem{}
Norberg P. \& the 2dFGRS team, 2002, MNRAS, 332, 827
%Norberg P. \& the 2dFGRS team, 2001c, MNRAS, astro-ph/0112043
%xi(L,eta)

\bibitem{}
O'Meara J.M. et al., 2001, ApJ, 552, 718 
%astro-ph/0011179

\bibitem{}
Peacock J.A. \& the 2dFGRS team, 2001, Nature, 410, 169

\bibitem{}
Peebles P. J. E. 1980, {\it  Large Scale Structure of the Universe},
Princeton University Press, Princeton.


\bibitem{}
Pen U., 1998, ApJ, 504, 601


\bibitem{}
Percival W.J. \& the 2dFGRS team, 2001, MNRAS, 327, 1297 (P01)
% astro-ph/0105252



\bibitem{}
Pierpaoli E., Scott D., White M., 2001, MNRAS, 325, 77


\bibitem{}
Pryke C. et al., 2002, ApJ, 568, 46
%2001, astro-ph/0104490 

\bibitem{}
Reiprich T.H., Boehringer H., 2002, 
ApJ, 567, 716, 
%astro-ph/0111285

\bibitem{}
Rhodes J., Refregier A.,  Groth E.J.,  
ApJ, submitted, 
astro-ph/0101213


\bibitem{}
Seljak U., 2001, 
MNRAS, submitted,
astro-ph/0111362

\bibitem{}
Seljak U.,  Zaldarriaga M. 1996, ApJ, 469, 437

\bibitem{}
Shepherd C.W. et al. 2001, ApJ, 560, 72
%astro-ph/0106250

\bibitem{}
Sigad Y., Branchini E., Dekel A., 1999, ApJ, 520, 24


\bibitem{}
Somerville R., Lemson G., Sigad Y., Dekel A.,
Colberg J., Kauffmann G., White S.D.M., 2001, MNRAS, 320, 289

\bibitem{}
Stompor R. et al., 2001, ApJ, 561, L7 
%astro-ph/0105062

\bibitem{}
Sutherland W. et al., 1999, MNRAS, 308, 289 


\bibitem{szalay01}
Szalay A.S. \& the SDSS team, 2001, ApJ, submitted, astro-ph/0107419

\bibitem{}
Tegmark M., Hamilton A.J.S.,  Xu Y., 2001, astro-ph/0111575

\bibitem{vw}
Van Waerbeke L., et al. 2001,  
AA, submitted, 
%submitted to A\&A, 
astro-ph/0101511 

\bibitem{}
Verde L. et al. \& the 2dFGRS team, MNRAS, accepted, astro-ph/0112161

\bibitem{}
Viana P.T.P., Nichol R.C., Liddle A.R.,  2002, 
ApJ, 569, L75
%astro-ph/0111394


\bibitem{}
Vittorio N.,  Silk J.,  1991, ApJ, 385, L9


\bibitem{}
Wang X., Tegmark M., Zaldarriaga M., 2001, 
PRD, accepted, astro-ph/0105091 (WTZ)


\bibitem{W98}
Webster M., Bridle S.L., Hobson M.P., Lasenby A.N., 
Lahav O., Rocha, G.,  1998, 
ApJ Lett, 509, L65

\bibitem{}
Zehavi I.,  et al., 2001, ApJ, accepted, astro-ph/0106476 


\end{thebibliography}
\end{document}